\documentclass[10pt,aps,prl,twocolumn,superscriptaddress,reprint]{revtex4-1}
\usepackage{xr}
\usepackage{graphicx}
\usepackage{float}
\usepackage{amsmath}
\usepackage{amssymb}
\usepackage{mathtools} 
\usepackage{color}
\usepackage{xcolor}
\usepackage[english]{babel}
\usepackage{silence}
\usepackage[T1]{fontenc}
\usepackage[utf8]{inputenc}

\usepackage[columnwise]{lineno}

\usepackage{hyperref}
\usepackage{physics}
\usepackage[caption=false]{subfig}

\newcommand{\spD}[1]{\fn{\tilde{\chi}_{_V}}{#1}}

\newcommand{\fn}[2]{\mathinner{#1\mathopen{\left(#2\right)}}}
\newcommand{\vect}[1]{{\bf #1}}

\newcommand{\epsTE}[1]{\fn{\varepsilon_e^\mathrm{TE}}{#1}}

\usepackage{soul}
\setstcolor{red}

\begin{document}

\title{Generating large disordered stealthy hyper\-uniform systems with ultra-high accuracy to determine their physical properties
}

\date{\today}
\author{Peter K. Morse}
\thanks{Corresponding author.}
\email{peter.k.morse@gmail.com}
\affiliation{Department of Chemistry, Princeton University, Princeton, NJ 08544}
\affiliation{Department of Physics, Princeton University, Princeton, NJ 08544}
\affiliation{Princeton Institute of Materials, Princeton University, Princeton, NJ 08544}
\author{Jaeuk Kim}
\affiliation{Department of Chemistry, Princeton University, Princeton, NJ 08544}
\affiliation{Department of Physics, Princeton University, Princeton, NJ 08544}
\affiliation{Princeton Institute of Materials, Princeton University, Princeton, NJ 08544}
\author{Paul J. Steinhardt}
\affiliation{Department of Physics, Princeton University, Princeton, NJ 08544}
\author{Salvatore Torquato}
\affiliation{Department of Chemistry, Princeton University, Princeton, NJ 08544}
\affiliation{Department of Physics, Princeton University, Princeton, NJ 08544}
\affiliation{Princeton Center for Theoretical Science, Princeton University, Princeton, NJ 08544}
\affiliation{Princeton Institute of Materials, Princeton University, Princeton, NJ 08544}
\affiliation{Program in Applied and Computational Mathematics, Princeton University, Princeton, NJ 08544}

\begin{abstract}
Hyper\-uniform many-particle systems are characterized by a structure factor $S({\mathbf{k}})$ that is precisely zero as $\abs{\mathbf{k}}\rightarrow0$;  and \textit{stealthy} hyper\-uniform systems have $S({\mathbf{k}})=0$ for the finite range ${0 < \abs{{\mathbf{k}}} \le K}$, called the ``exclusion region.''
Through a process of \textit{collective-coordinate} optimization, {\it energy-minimizing} disordered stealthy hyper\-uniform systems of moderate size have been made to high accuracy, and their novel physical properties have shown great promise. However, minimizing $S(\mathbf{k})$ in the exclusion region is computationally intensive as the system size becomes large.
In this Letter, we present an improved methodology to generate such states using double-double precision calculations on GPUs that reduces the deviations from zero within the exclusion region by a factor of approximately $10^{30}$ for systems sizes more than an order of magnitude larger.
We further show that this ultra-high accuracy is required to draw conclusions 
about their corresponding characteristics, such as the nature of the associated energy landscape and the presence or absence of Anderson localization, which might be masked, even when deviations are relatively small.

\end{abstract}

\maketitle

\textit{ Introduction.---}
Disordered hyper\-uniform systems are isotropic like a glass but have long-range correlations that result in the anomalous suppression of density fluctuations over large distances~\cite{torquato_local_2003}.  This feature endows them with unique physical properties not possible in ordered (periodic or quasiperiodic) systems~\cite{wu_effective_2017, salvalaglio_hyperuniform_2020, yu_engineered_2021, yu_evolving_2023}.  They have been shown to be surprisingly ubiquitous; they include perfect glasses, fermionic point processes, disordered jammed particle packings,  quantum states, certain plasmas, galaxy distributions, eigenvalues of random matrices, and a myriad of other examples (see Ref.~\cite{torquato_hyperuniform_2018} and references therein).

A defining feature of a disordered hyper\-uniform system in $d$-dimensional Euclidean space $\mathbb{R}^d$, aside from its isotropy, is that the structure factor $S({\mathbf{k}})$ vanishes as the wavenumber $k\equiv|{\vect{k}}|$ tends to zero~\cite{torquato_local_2003, torquato_hyperuniform_2018}.
An important subclass is the {\it stealthy} disordered hyper\-uniform system in which $S({\mathbf{k}}) = 0$ for a finite range of wavenumbers ${0 < k \leq K}$. This range defines the {\it exclusion region} in Fourier space where no single-scattering events can occur~\cite{uche_constraints_2004, uche_collective_2006, batten_classical_2008, torquato_ensemble_2015}, leading to them being theoretically analyzed as $d$-dimensional 
``hard-sphere fluids'' in Fourier space ~\cite{torquato_ensemble_2015}.
Disordered stealthy hyper\-uniform systems stand out among all non-stealthy hyper\-uniform ones because they anomalously suppress density fluctuations not only at infinite wavelengths but down to intermediate wavelengths~\cite{torquato_hyperuniform_2018}. Moreover,  ``holes'' in any disordered stealthy hyper\-uniform point pattern are strictly bounded with a well-defined maximal size in the thermodynamic limit~\cite{zhang_can_2017, ghosh_generalized_2018, torquato_hyperuniform_2018}. All of these remarkable attributes are responsible for  the novel wave, transport, and mechanical properties of disordered stealthy hyper\-uniform systems~\cite{torquato_hyperuniform_2018}.  For example, a scheme that mapped stealthy point patterns onto disordered dielectric networks resulted in the first moderately-sized amorphous photonic solid samples with complete photonic band gaps comparable to those in periodic networks with the advantage that the band gaps are isotropic~\cite{florescu_designer_2009}. Since this work, disordered stealthy systems have been extensively fabricated~\cite{man_isotropic_2013, leseur_highdensity_2016, zhang_transport_2016, wu_effective_2017, froufe-perez_band_2017, chen_designing_2018, gorsky_engineered_2019, sheremet_absorption_2020, kim_multifunctional_2020, tavakoli_65_2022, gkantzounis_hyperuniform_2017, romero-garcia_stealth_2019, rohfritsch_impact_2020, romero-garcia_wave_2021, yu_engineered_2021, cheron_wave_2022,  klatt_wave_2022} and studied computationally especially due to their novel optical~\cite{leseur_highdensity_2016, wu_effective_2017, chen_designing_2018, sheremet_absorption_2020, kim_multifunctional_2020, tavakoli_65_2022}, photonic~\cite{man_isotropic_2013, froufe-perez_band_2017, gorsky_engineered_2019, yu_engineered_2021, klatt_wave_2022}, electronic~\cite{torquato_multifunctional_2018, granchi_nearfield_2022}, phononic~\cite{gkantzounis_hyperuniform_2017, romero-garcia_stealth_2019, rohfritsch_impact_2020, romero-garcia_wave_2021}, and transport~\cite{zhang_transport_2016, cheron_wave_2022} properties.

Since hyper\-uniformity is a large-scale property of a system, it is imperative to be able to generate sample sizes much larger than those presently possible~\cite{zhang_realizable_2020} in order to ascertain whether their novel properties persist as the sample size becomes large.
Indeed, this question was recently explored in the context of photonics, where numerical evidence was presented to show that among various disordered nonhyper\-uniform and hyper\-uniform network dielectric solids, only certain stealthy hyper\-uniform ones may form bandgaps in the thermodynamic limit~\cite{klatt_wave_2022}. Importantly, the bandgaps for networks with near stealthiness (small but positive $S({\bf k})$ within the exclusion region) eventually closed as the system size grew with a rate inversely proportional to the {\it distance to stealthiness} $S_\mathrm{max}$,
which we define in the present work as
\begin{equation}
S_\mathrm{max} \equiv \max_{0<k \leq K} S(\mathbf{k}).
\label{eq:smax}
\end{equation}
The work reported in Ref.~\cite{klatt_wave_2022} emphasizes the importance of computationally creating appreciably larger disordered stealthy hyper\-uniform systems, but ones with the smallest value of  $S_\mathrm{max}$
within the exclusion region.

It is a remarkable fact that disordered stealthy hyper\-uniform point patterns are highly degenerate ground states for certain non-trivial oscillatory pair potentials (see Ref. ~\cite{torquato_ensemble_2015} and references therein).
Such ground states have been generated using a collective-coordinate energy optimization scheme~\cite{fan_constraints_1991, uche_constraints_2004, batten_classical_2008, zhang_ground_2015}.
As such, they inevitably have a small positive residual $S(\mathbf{k})$ in the exclusion region, i.e., small $S_\mathrm{max}$, due to the level of precision with which they were prepared.
In spatial dimensions $d=1$-$3$, the largest simulated disordered stealthy hyper\-uniform systems have had $N=10^3$, $10^4$, and $8\times10^3$ particles, respectively~\cite{torquato_local_2021}, and the highest accuracy achieved has been $S_\mathrm{max} \approx 10^{-22}$~\cite{zhang_ground_2015}. 
In order to extrapolate reliably to the thermodynamic limit, significantly larger systems with a small $S_\mathrm{max}$ are required.
However, the computational cost of the standard collective-coordinate minimization schemes scales as at least $\mathcal{O}(N^2)$ (as derived below), making it nearly impossible to reach the requirements  without fundamentally modifying the techniques used.

In this Letter, we show that the collective-coordinate minimization is highly parallelizable and requires very little memory access, making it ideally suited for a Graphical Processing Unit (GPU) based algorithm. The high degree of parallelization afforded by GPUs allows us to create systems $20$ times larger than previous best efforts~\cite{torquato_local_2021} in comparable time and with $S_\mathrm{max}$ reduced by $30$ orders of magnitude, i.e. $S_\mathrm{max}\approx10^{-51}$ in dimensions $d=1$-$4$. This enables the exploration of characteristics of the energy landscape associated with the stealthy hyper\-uniform  potential and lays the groundwork for extrapolations to both the thermodynamic and high-dimensional ($d\to\infty$) limit. While the techniques used are independent of dimension, in this Letter we focus on $d=2$ with system size $N=2\times10^6$ as a trial case for its ease of representation, and we use random initial conditions---corresponding to a high temperature---to show the generic character of these results. These systems can then be  compared to the low-temperature theory of stealthy systems~\cite{torquato_ensemble_2015} via the $\tau$ order metric, which characterizes their degree of translational disorder.

Moreover, we demonstrate the importance of creating states
with the smallest distance to stealthiness by showing that values of $S_\mathrm{max}$ which were deemed relatively small in the past can still vastly degrade certain desired physical properties of a stealthy hyper\-uniform material, in addition to those reported in Ref.~\cite{klatt_wave_2022}. For each application, it is then necessary to ask what level of $S_\mathrm{max}$ is sufficient. In particular, we show that for two-dimensional two-phase media derived from stealthy hyper\-uniform point patterns, a full transparency interval for light propagation in which there is no Anderson localization depends on having $S_\mathrm{max}$ precisely zero over the excluded region. Thus being able to generate stealthy hyper\-uniform systems over an exponentially wide range of $S_\mathrm{max}$ is essential for extrapolating to the ideal condition in the thermodynamic limit. This degree of accuracy is also necessary in determining key properties of the ground state manifold of the stealthy hyper\-uniform potential---namely its connectedness and dimension---especially near critical points. 
 
{\it Definitions.} The collective-coordinate method for generating disordered stealthy hyper\-uniform point patterns is based on finding the ground state for a system of point-particles whose total potential energy is the sum of pair-wise potentials $v(\vect{r})$, where $v(\vect{r})$  is bounded and integrable such that its Fourier transform $\tilde{v}(\vect{k})$ exists and is a positive function with compact support over the interval $0 < |\vect{k}| \leq K$.
Given a set of $N$ points at positions $\mathbf{r}_j$ within a periodic box $F$ of volume $v_F$ in $\mathbb{R}^d$, The total potential energy $\Phi(\mathbf{r}^N)$ has the Fourier representation~\cite{torquato_ensemble_2015}:
\begin{equation}
\Phi(\mathbf{r}^N) = \frac{N}{2v_F} \qty[ \sum_{0<k\leq K} \tilde{v}(\mathbf{k})\fn{\mathcal{S}}{\vect{k}} - \sum_{0<k\leq K} \tilde{v}(\mathbf{k}) ],
\label{eq:energy}
\end{equation}
where $\mathcal{S}(\vect{k})$ is the structure factor of a \emph{single} configuration defined by
\begin{equation}    \label{eq:structurefactor}
\mathcal{S}(\mathbf{k}) = \frac{|\tilde{n}(\mathbf{k})|^2}{N},
\end{equation}
where $\vect{k}$ is a non-zero reciprocal lattice vector of $F$, and $\tilde{n}(\mathbf{k})$ is the complex collective density variable given by
\begin{equation}
\tilde{n}(\mathbf{k}) = \sum_{j=1}^N \exp(-i\mathbf{k}\cdot\mathbf{r}_j).
\label{eq:ntilde}
\end{equation}
Note that the second term in Eq. \eqref{eq:energy} is independent of structure so that one can drop this term in practice. The potential $\Phi$ is then bounded from below by $\Phi=0$, which defines the ground state manifold within which all states are stealthy hyper\-uniform. Because $\tilde{v}(\mathbf{k})$ is strictly positive, it does not modify the definition of the ground state manifold, though it does provide a weight function that will funnel any dynamic or thermal processes on the $\Phi$ landscape towards specific ground states.

\begin{figure*}[ht!]
\includegraphics[width=\textwidth]{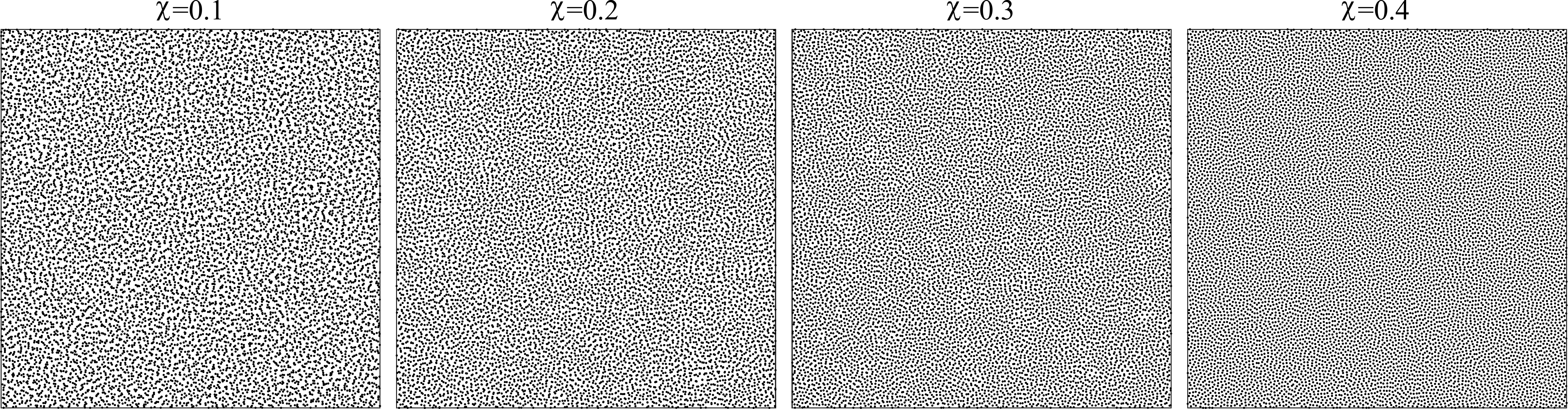}
\caption{Stealthy hyper\-uniform point patterns of $N=2\times10^6$ particles with $\chi=0.1$, $0.2$, $0.3$, and $0.4$ at the same number density $\rho\equiv N/v_F$. Each figure is only showing $1/16^\mathrm{th}$ of all the data so as to allow for better visualizations. Full configuration data for 5 iterations of each $\chi$ value are included in the Supplemental Material~\cite{suppData}.}
\label{fig:2dvis}
\end{figure*}

Because the periodic hypercubic box has a finite side length $L$, all $k$-vectors are integer linear combinations of the   minimum $k$-vectors with length $k_\mathrm{min}= 2 \pi/L$, so one can count the number for which $0<k\le K$. Because the structure factor has inversion symmetry, i.e. $S(\mathbf{k})=S(-\mathbf{k})$, if there were $(2M+1)$ vectors for which $0<k\le K$, then only $M$ are independent. We thus state that there are $M$ independent constraints on the system. Given that there are a total of $d(N-1)$ degrees of freedom in a $d$-dimensional point pattern, the fraction of degrees of freedom that are constrained is
\begin{equation}
\chi = \frac{M}{d(N-1)}.
\end{equation}
Because $M$ is a monotonically increasing function of $K$, $\chi$ additionally acts as a measure of the size of the exclusion region. It has been shown that ground states with $\chi \ge 1/2$ in dimensions for $d \le 4$ always form ordered hyper\-uniform structures, so we restrict ourselves in this study of $d=2$ patterns  to $\chi < 1/2$, for which
the ordered structures within the ground state manifold are of measure zero~\cite{torquato_ensemble_2015}.
Results for one-dimensional as well as three- and higher-dimensional cases will be reported in a future study.

The energy landscape for the stealthy hyper\-uniform potential in the disordered regime ($\chi < 1/2$ or $\chi < 1/3$ in $d=1$) is remarkably simple. Previous studies have shown that the \textit{inherent-structure state} for any initial condition---defined as the nearest local energy minimum accessible without a barrier crossing~\cite{stillinger_packing_1984}---is itself one of the many degenerate global minima~\cite{uche_constraints_2004, uche_collective_2006, batten_classical_2008, zhang_ground_2015}. Thus, a simple quench, achieved numerically via standard local energy minimization techniques, will generically produce a ground state. 
While previous studies found this to be true to an assigned (but arbitrary) threshold of $\Phi < 10^{-22}$~\cite{zhang_ground_2015}, this assumption will be verified to greater accuracy in this work.

{\it Simulations.} Generating disordered stealthy hyper\-uniform systems reduces to finding true ground states of the total potential energy $\Phi$ in  Eq.~\eqref{eq:energy}.  When $\chi < 1/2$, the spectrum of disordered ground states is highly degenerate with an extent that depends on the value of $\chi$, becoming infinitely degenerate in the thermodynamic limit~\cite{torquato_ensemble_2015}.   In the collective-coordinate optimization procedure, both the initial conditions and the choice of $\tilde{v}(\mathbf{k})$ affect the particular ground state obtained~\cite{zhang_ground_2015}, though the methods used are identical. 
For the purposes of illustration, we use random initial conditions defined by a Poisson point process and the simplest form of $\tilde{v}(\mathbf{k})$. Here $\tilde{v}(\mathbf{k})=\varepsilon_0L^d\Theta(K-k
)$, where $\Theta(x)$ is the Heaviside function and $\varepsilon_0$ sets the units of energy. 

In order to minimize Eq.~\eqref{eq:energy}, we need its value and the gradient 
\begin{equation}
\mathbf{F}_j = -\nabla_j \Phi(\mathbf{r}^N) = \frac{1}{v_F}\sum_{0<k\leq K}\mathbf{k}\tilde{v}(\mathbf{k})\mathrm{Im}[\tilde{n}(\mathbf{k})\exp(i\mathbf{k}\cdot\mathbf{r}_j)],
\label{eq:force}
\end{equation}
which can be used as inputs into a minimizer. Here we use FIRE minimization~\cite{bitzek_structural_2006}, because it requires fewer computations and converges faster than other comparable minimizers.

By far, the most computationally expensive process of the minimization is the calculation of the gradient, which contains two sums. First, it is necessary to calculate all $M$ independent values of $\tilde{n}(\mathbf{k})$, each of which contains $N$ terms. Then, one must calculate the gradient for all $N$ particles, each of which contains a sum over $M$ terms. There are thus 2 calculations---each with ${MN=\chi N(N-1)d}$ terms---which need to be recalculated at every step of the minimization, setting the minimum time scale for minimization as $\mathcal{O}(N^2)$. The substantial numerical speedup comes from the realization that both Eq.~\eqref{eq:ntilde} and Eq.~\eqref{eq:force} are simple parallel reduction sums, which are ideally suited for GPUs. 
To implement these quenches, we use the GPU-based packing software pyCudaPack~\cite{charbonneau_universal_2012, morse_geometric_2014, charbonneau_jamming_2015, morse_geometric_2016, morse_echoes_2017, arceri_vibrational_2020, morse_direct_2021}, because of its highly-optimized modular implementation of a variety of pair interactions~\cite{charbonneau_universal_2012, morse_geometric_2014, arceri_vibrational_2020, morse_direct_2021} using various minimizers~\cite{morse_geometric_2014} with tunable precision using double-doubles~\cite{charbonneau_jamming_2015, morse_geometric_2016, morse_echoes_2017}. Minimizations proceed until $\Phi$ is at the minimum value attainable within machine precision, which for $N=2\times10^6$ in unitless form is ${\Phi/\varepsilon_0\approx 10^{-46}}$. 

It is well-established that short- and intermediate-range order increase with the size of the exclusion region between $\chi=0$ and $1/2$ in $d=2$~\cite{uche_constraints_2004, uche_collective_2006, batten_classical_2008, zhang_ground_2015, torquato_ensemble_2015}.
A positive order metric that measures the degree to which translational order increases with $\chi$ across length scales is~\cite{torquato_ensemble_2015}
\begin{equation}   \label{eq:tau}
\tau \equiv \frac{1}{D^d} \int_{\mathbb{R}^d} \fn{h^2}{r} \dd{\vect{r}} 
= \frac{1}{(2\pi D)^d\rho^2}\int_{\mathbb{R}^d} [S(\mathbf{k})-1]^2 \dd{\vect{k}},
\end{equation}
where $\rho$ is the number density in the thermodynamic limit, $\fn{h}{r}$ is the total correlation function~\cite{totalCorrelation}, $\fn{S}{\vect{k}}$ is the ensemble average of Eq. \eqref{eq:structurefactor} in the thermodynamic limit, and $D$ is a characteristic length scale, which we take to be $D=K^{-1}$.
For an ideal gas (spatially uncorrelated Poisson point process), $\tau=0$ because $\fn{h}{r}=0$ for all $r$. 
Thus, a deviation of $\tau$ from zero measures translational order with respect to the fully uncorrelated case.
While $\tau$ diverges for any perfect crystal and quasicrystal in the infinite-system limit, its rate of growth as a function of system size still provides a useful measure of the translational order in such ordered systems~\cite{torquato_uncovering_2018, torquato_hidden_2019}.

\begin{figure}[ht]
\includegraphics[width=\columnwidth]{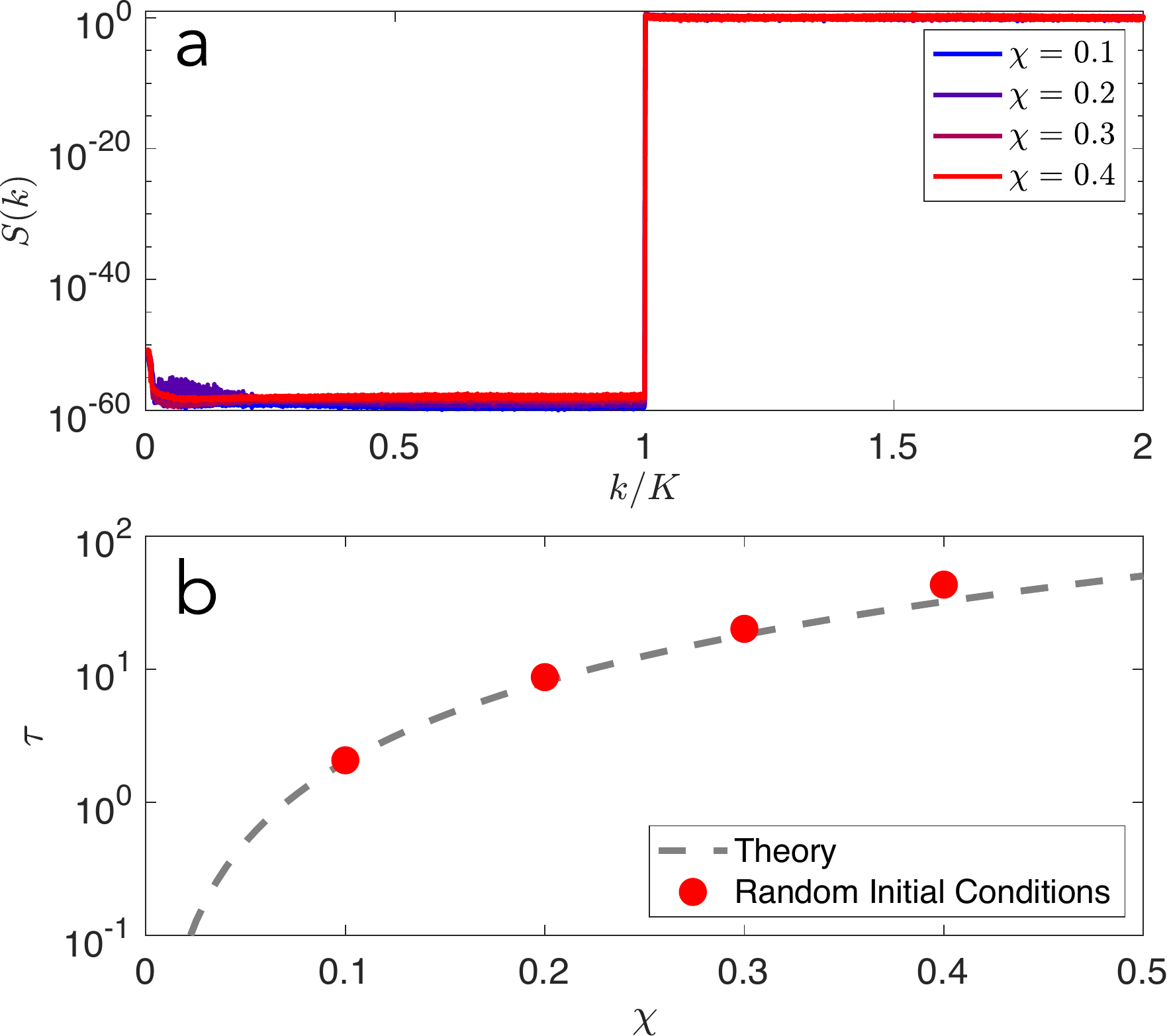}
\caption{(a) The distance to stealthiness is $S_\mathrm{max} \approx 10^{-51}$ as stated, which is below the threshold for what can be distinguished from zero to within machine precision.
(b) The order metric $\tau$, defined by \eqref{eq:tau}, plotted as a function of $\chi$ for the simulation data (red circles) compared to the theoretical curve from the theory of entropically favored states~\cite{torquato_ensemble_2015}.}
\label{fig:sofk2d}
\end{figure}

{\it Results and Discussion.---} We create five disordered
stealthy hyper\-uniform systems in $d=2$
at each value of $\chi=0.1$, $0.2$, $0.3$, and $0.4$ in order to demonstrate the ultra-high accuracy of our GPU-based algorithm. Our results have both high-precision---in that they are done using double-doubles--- and high accuracy, as measured by exceptionally small values of $S_\mathrm{max}$. Figure~\ref{fig:2dvis} shows one system at each value of $\chi$, though for easier visualization, only $1/16^\mathrm{th}$ of the data is shown. Full configuration data for all 20 systems are given in the Supplemental Material~\cite{suppData}. To demonstrate the level of accuracy, we plot the structure factor on a logarithmic scale (Fig.~\ref{fig:sofk2d}(a)).
The $k$-dependence of $S(0<k\leq K)$, and $S_\mathrm{max}$ in particular, depend explicitly on the form of $\tilde{v}(\mathbf{k})$, with our particular choice yielding $S_\mathrm{max}=S(k_\mathrm{min})$.

While $S(k>K)$ has been shown to depend strongly on initial temperature (see Ref.~\cite{zhang_ground_2015} and the Supplemental Material~\cite{suppMat}), in Fig.~\ref{fig:sofk2d}(b) we find that the theoretical curve for $\tau$ derived for low-temperature initial states departs by relatively small amounts compared to the data obtained from high-temperature states (i.e. random initial conditions). These results imply that while the respective pointwise behaviors of the pair correlation function and $S(\mathbf{k})$ may sometimes differ for certain small ranges of their arguments, the integrated measure of order across length scales for stealthy hyperuniform states from low-$T$ and high-$T$ initial states, as measured by $\tau$, are essentially the same. Whether this remains true in other dimensions is a subject for future work.

\begin{figure}[h]
\includegraphics[width=\columnwidth]{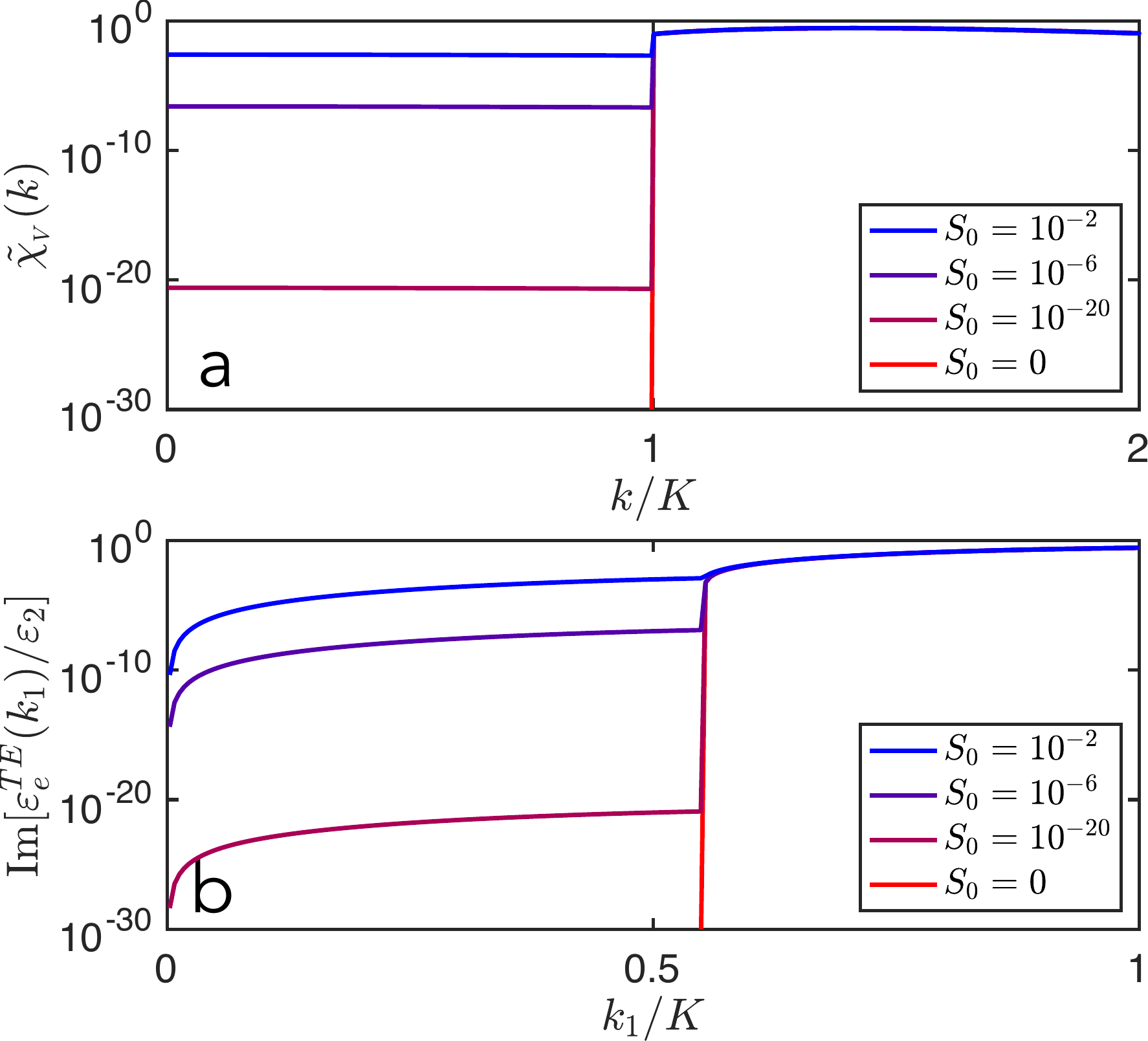}
\caption{Two-dimensional media consisting of packings of identical disks of dielectric constant $\varepsilon_2=1$ in a matrix of dielectric constant $\epsilon_1=11.6$.
The nonoverlapping disks have radius $a$, and their packing fraction is $\phi_2=0.112$.
The disk centers have $\fn{S}{k}=S_\mathrm{max}$ in the exclusion region ($k\leq K)$.
(a) Semi-log plot of the spectral density $\fn{\tilde{\chi}_{_V}}{k}$ as a function of a dimensionless wavenumber $k/K$. (b) Semi-log plot of the imaginary part of the effective dynamic dielectric constant $\Im[\fn{\varepsilon_e^\mathrm{TE}}{k_1}]$ for transverse electric (TE) polarization as a function of the dimensionless incident wavenumber $k_1/K$ in the matrix phase $1$. }
\label{fig:equilum}
\end{figure}

We now vividly demonstrate importance of the small distance to stealthiness in stealthy hyper\-uniform configurations  by mapping  onto two-phase dielectric media and quantifying the degree of the attenuation of electromagnetic wave propagation through them.
Specifically, we map 2D stealthy hyperuniform point patterns that are exactly stealthy (i.e., $S_\mathrm{max}=0$) in the thermodynamic limit via theoretical methods described in Ref.~\cite{torquato_ensemble_2015} into a distribution of disks (phase 2) of dielectric constant $\varepsilon_2 = 1$ in a matrix (phase 1) of dielectric constant $\varepsilon_1=11.6$ by circumscribing each point with identical disks of radius $a$ without overlap~\cite{torquato_disordered_2016, zhang_transport_2016}.
By a similar analysis, we allow $S_\mathrm{max}$ to be a free parameter.
The resulting area fraction covered by disks is $\phi_2=0.112$.
The \emph{spectral density} $\spD{k}$ of these disks is computed from the formula~\cite{torquato_disordered_2016}, $\spD{k} = 4\pi \phi_2 [J_1(ka)/k]^2 \fn{S}{k}$, where $J_1(x)$ is the Bessel function of the first kind of order 1.
The resulting $\spD{k}$ is valid in the thermodynamic limit and inherits the distance to stealthiness $S_\mathrm{max}$ of the original point patterns, in which $S_\mathrm{max}$ is either exactly zero or a specified positive value~\cite{Sk}; see Fig.~\ref{fig:equilum}(a).
We obtain the effective dynamic dielectric constant $\epsTE{k_1}$ of such two-phase media for incident light of transverse electric (TE) polarization and wavenumber $k_1$ by using a nonlocal strong-contrast approximation~(see Eq. (73) for $d=2$ in~\cite{torquato_nonlocal_2021}) that depends on $\fn{\tilde{\chi}_{_V}}{k}$, which accurately captures multiple scattering effects to all orders beyond the quasistatic regime (i.e., $0\leq k_1 \xi \lesssim 1$, where $\xi$ is a characteristic inhomogeneity length scale).

The key property of interest in this system is the imaginary part of $\epsTE{k_1}$ which measures the degree to which the media effectively attenuate light, as shown in Fig.~\ref{fig:equilum}(b).
Importantly, perfect stealthy hyper\-uniform media with $S_\mathrm{max}=0$ exhibit full transparency (i.e., $\Im[\fn{\varepsilon_e^\mathrm{TE}}{k_1}]=0$) up to a finite wavenumber in the thermodynamic limit, implying the absence of Anderson localization~\cite{mcgurn_anderson_1993, aegerter_coherent_2009, sgrignuoli_subdiffusive_2022}.
By contrast, a stealthy hyper\-uniform medium with a moderate distance to stealthiness $(0 \ll S_\mathrm{max} < 1)$ has a positive $\Im[\epsTE{k_1}]$ proportional to $S_\mathrm{max}$ up to $k_1=\mathcal{O}(K)$, implying that transmittance through this medium is increasingly suppressed with a larger sample size or a larger $S_\mathrm{max}$, and that Anderson localization likely emerges.
Therefore, to better understand  the insulator-conductor transition as the system size increases, it is essential to generate much larger sample sizes with a small $S_\mathrm{max}$ via the techniques described in this paper.

\emph{Conclusions}---
Through the use of GPU minimizations with double-double precision, we have been able to dramatically increase the size and reduce the distance to stealthiness
of stealthy hyper\-uniform point patterns.
The ability to create disordered stealthy hyper\-uniform systems of both large sizes and ultra-small $S_\mathrm{max}$ is imperative to study their novel properties.
While the requirement for large sample sizes has been recognized, the need for a  small $S_\mathrm{max}$ has gone largely unnoticed due to the inability to create large samples.
Indeed, most previous studies were based on relatively small sample sizes ($100 \lesssim N \lesssim 1000$) with $10^{-10} \lesssim S_\mathrm{max} \lesssim 10^{-5}$.
Using a capacity to create much larger systems with very small values of $S_\mathrm{max}$, we can shed light 
on  the emerging novel properties of stealthy hyper\-uniform systems that
depend on having a small distance to stealthiness.
One such open question is whether isotropic photonic (or phononic) band gaps exist in two-phase systems derived from stealthy hyper\-uniform point patterns in the thermodynamic limit~\cite{klatt_wave_2022}.
In addition, while disordered stealthy hyper\-uniform systems of moderate sizes are known to be transparent up to a finite $k$ for electromagnetic~\cite{leseur_highdensity_2016, froufe-perez_band_2017,torquato_nonlocal_2021} and elastic waves~\cite{rohfritsch_impact_2020, romero-garcia_wave_2021, kim_multifunctional_2020}, it is not known whether this transparency interval persists in the thermodynamic limit without Anderson localization.
Our theoretical analysis [summarized in  Fig.~\ref{fig:equilum}(b)] has shown that this transparency property can increasingly degrade as $S_\mathrm{max}$ and the system size are made larger, stressing the importance of making $S_\mathrm{max}$ for a given system as small as possible, which is now achieved with ultra-high accuracy via our improved numerical methodology.
Samples of much larger size with vastly smaller $S_\mathrm{max}$, heretofore not possible, can now be potentially  fabricated by combining our designs with photolithographic and 3D printing techniques~\cite{tumbleston_continuous_2015, shirazi_review_2015, zhao_assembly_2018} in order to explore the presence or absence of Anderson localization~\cite{mcgurn_anderson_1993, aegerter_coherent_2009, wiersma_disordered_2013} in disordered stealthy hyperuniform systems as function of sample size.

These new high precision results also lay the groundwork to further investigate the characteristics of the energy landscape of the stealthy hyper\-uniform potential.
Indeed, we now have abundant evidence that local minima do not exist (or are extremely rare) for $\chi < 1/2$, implying that states with high values of $S_\mathrm{max}$ found in previous studies are not locally stable; rather, they are the result of incomplete minimization, making them excited states with an effective temperature and a modified isothermal compressibility~\cite{torquato_ensemble_2015}. Future work will aim to understand the nature of the true ground state manifold, particularly its connectivity, the topology near crystalline states, and the topological change that occurs when $\chi=1/2$ (especially as a function of dimension $d$) and how these landscape properties modify the novel physical properties of disordered stealthy hyper\-uniform systems. 
Moreover, with significantly larger systems, one will be able to probe higher-order correlation functions beyond $\tau$ in determining the structural order in stealthy hyper\-uniform systems~\cite{torquato_local_2021}.
Additionally, while we have focused here on isotropic exclusion regions, it is straightforward to apply our ultra-high accuracy method to create larger stealthy hyperuniform systems with anisotropic exclusion regions, and thus statistically anisotropic ground states~\cite{martis_exotic_2013, torquato_hyperuniformity_2016}.

\begin{acknowledgments}
The Research was sponsored by the Army Research Office and was accomplished under Cooperative Agreement Number W911NF-22-2-0103. Simulations were performed on computational resources managed and supported by the Princeton Institute for Computational Science and Engineering (PICSciE). P.K.M. would like to thank Eric Corwin for discussions related to pyCudaPack. 
\end{acknowledgments}

\bibliographystyle{apsrev4-1}
\bibliography{hyperuniform, programs}

\begin{thebibliography}{60}
\bibitem [{tot()}]{totalCorrelation}%
  \BibitemOpen
  \bibinfo {note} {The total correlation function $h(r)$ is related to the pair
  correlation function $g_2(r)$ as $h(r)\equiv g_2(r)-1$.}\BibitemShut {Stop}%
\end{thebibliography}

\end{document}


\renewcommand{\theequation}{S\arabic{equation}}
\renewcommand{\thefigure}{S\arabic{figure}}
\renewcommand{\thetable}{S\arabic{table}}

\title{Supplementary Material - Generating large disordered stealthy hyperuniform systems with ultra-high accuracy to determine their physical properties}
\date{\today}
\author{Peter K. Morse}
\thanks{Corresponding author.}
\email{peter.k.morse@gmail.com}
\affiliation{Department of Chemistry, Princeton University, Princeton, NJ 08544}
\affiliation{Department of Physics, Princeton University, Princeton, NJ 08544}
\affiliation{Princeton Institute of Materials, Princeton University, Princeton, NJ 08544}
\author{Jaeuk Kim}
\affiliation{Department of Chemistry, Princeton University, Princeton, NJ 08544}
\affiliation{Department of Physics, Princeton University, Princeton, NJ 08544}
\affiliation{Princeton Institute of Materials, Princeton University, Princeton, NJ 08544}
\author{Paul J. Steinhardt}
\affiliation{Department of Physics, Princeton University, Princeton, NJ 08544}
\author{Salvatore Torquato}
\affiliation{Department of Chemistry, Princeton University, Princeton, NJ 08544}
\affiliation{Department of Physics, Princeton University, Princeton, NJ 08544}
\affiliation{Princeton Center for Theoretical Science, Princeton University, Princeton, NJ 08544}
\affiliation{Princeton Institute of Materials, Princeton University, Princeton, NJ 08544}
\affiliation{Program in Applied and Computational Mathematics, Princeton University, Princeton, NJ 08544}

\maketitle

\section{Pair correlation function and structure factor}

\begin{figure}[htp!]
\includegraphics[width=\columnwidth]{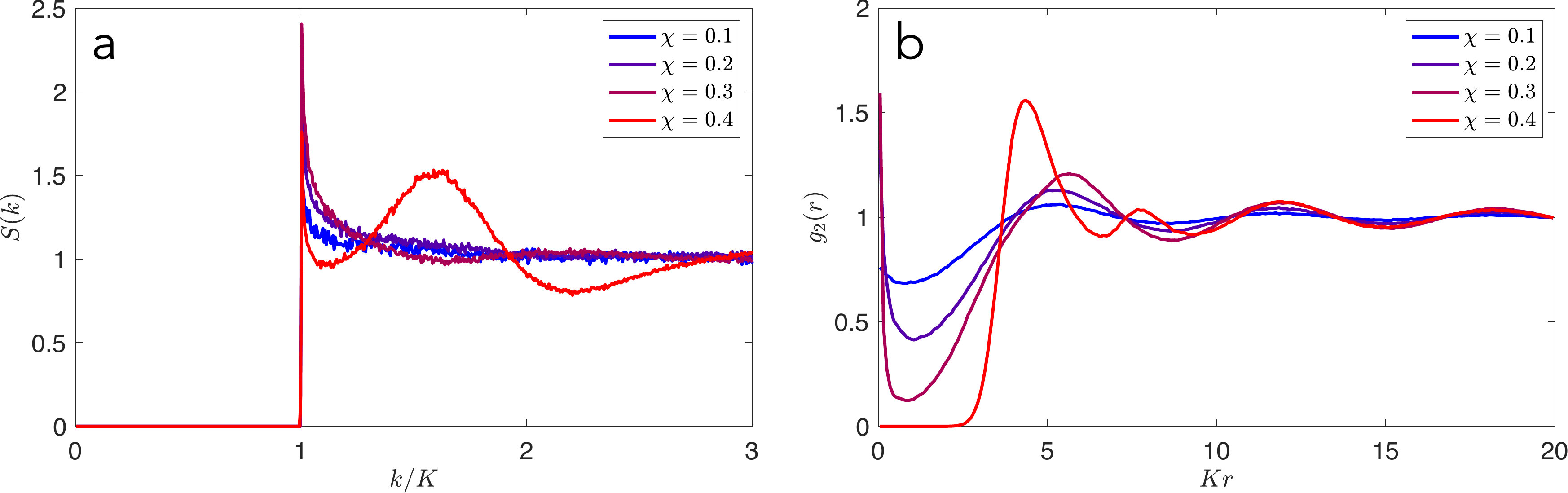}
\caption{(a) Data from Fig. 2 in the main text plotted on a linear scale. (b) The pair correlation function $g_2(r)$ for the same set of data. Note here the exclusion region forming for low $r$ as $\chi$ increases. Length scales are set by the choice $K=1$. Both (a) and (b) are consistent with Fig. 1 of Ref.~\cite{zhang_ground_2015}.}
\label{fig:g2skChi}
\end{figure}

In Fig. 2 of the main text, the structure factor was shown on a semi-log scale in order to demonstrate the level of precision achieved in creating stealthy hyperuniform systems with a minimal $S_\mathrm{max}$. Here, in Fig.~\ref{fig:g2skChi}a, it is shown on the more familiar linear scale. Also included is a plot of the pair correlation function $g_2(r)$; see Fig.~\ref{fig:g2skChi}(b). As noted throughout the text, this data is taken from random initial conditions and thus does not follow the predictions for entropically favored states shown in~\cite{torquato_ensemble_2015}. Instead, it should be compared to Fig. 1 of Ref.~\cite{zhang_ground_2015}, with which it is entirely consistent.



\bibliography{hyperuniform, programs}